\documentclass[preprint2,longabstract]{aastex}

\slugcomment{To appear in the Astrophysical Journal Letters}

\shorttitle{[NII] in the Carina Nebula}
\shortauthors{Oberst et al.}

\begin{document}


\title{Detection of the 205 $\mu$m [NII] Line from the Carina Nebula}


\author{T.E. Oberst, S.C. Parshley, G.J. Stacey, and T. Nikola} 
\affil{Cornell University, Ithaca, NY 14853}
\author{A.\ L\"{o}hr, J.I.Harnett, N.F.H. Tothill, A.P. Lane, and A.A. Stark}
\affil{Harvard-Smithsonian Center for Astrophysics, Cambridge, MA 02138}
\and
\author{C.E. Tucker}
\affil{University of Wales, Cardiff, Cardiff, UK CF24 3YB}



\begin{abstract}
We report the first detection of the 205 $\mu$m $^{3}P_{1}\rightarrow ^{3}P_{0}$ [NII] line from a ground-based observatory using a direct detection spectrometer.  The line was detected from the Carina star formation region using the South Pole Imaging Fabry-Perot Interferometer (SPIFI) on the Antarctic Submillimeter Telescope and Remote Observatory (AST/RO) at South Pole.  The [NII] 205 $\mu$m line strength indicates a low-density (n $\sim 32 \rm~{cm}^{-3}$) ionized medium, similar to the low-density ionized halo reported previously in its [OIII] 52 and 88 $\mu$m line emission.   When compared with the ISO [CII] observations of this region, we find that 27$\%$ of the [CII] line emission arises from this low-density ionized gas, but the large majority ($\sim$ 73$\%$) of the observed [CII] line emission arises from the neutral interstellar medium.  This result supports and underpins prior conclusions that most of the observed [CII] 158 $\mu$m line emission from Galactic and extragalactic sources arises from the warm, dense photodissociated surfaces of molecular clouds.  The detection of the [NII] line demonstrates the utility of Antarctic sites for THz spectroscopy.\end{abstract}





\section{Introduction}

The far-infrared (30 $\mu$m $\la \lambda \la$ 350 $\mu$m) contains a wide variety of fine-structure lines arising from the ground-state configurations of astrophysically abundant atoms and ions.  Among the brightest of these are the 205 $\mu$m and 122 $\mu$m [NII] and 158 $\mu$m [CII] lines.  

Since the ionization potential of nitrogen is 14.53 eV,  N$^{+}$ is only found in HII regions.  The ground-state $^{3}$P term of the N$^{+}$ ion is split by the spin-orbit interaction into the three $^{3}$P$_{2,1,0}$ levels from which the 121.898 $\mu$m ($^{3}$P$_{2} \rightarrow ^{3}$P$_{1}$) and 205.178 $\mu$m ($^{3}$P$_{1} \rightarrow ^{3}$P$_{0}$) lines arise.  For electron temperatures of 8000 K, the critical densities are 293 and 44 cm$^{-3}$ for the 122 and 205 $\mu$m lines, respectively, so that the line ratio is an excellent density probe for low-density ionized gas.  Both of the [NII] lines were strongly detected from the Milky Way by the COBE FIRAS experiment.  After the 158 $\mu$m [CII] line, these lines are by far the brightest lines in the 100 $\mu$m to 1 mm FIRAS bandpass, with line luminosities approximately 0.045$\%$ and 0.03$\%$ of the total Milky Way far-IR luminosity for the 122 $\mu$m and 205 $\mu$m lines, respectively (Wright et al. 1991).  ISO observations show that the [NII] 122 $\mu$m line is very bright in external galaxies as well, especially in the disks of normal galaxies (e.g., M83 and M51, Kramer et al. 2005).

Despite the strength of the lines, the [NII] lines were the last of the bright far-infrared fine-structure lines observed from NASA's airborne observatories in the 1979-1995 period.  This is because the 122 $\mu$m line lies very close to a strong telluric water vapor absorption feature, so that telluric transmission is poor ($\sim$ 30$\%$) even at high aircraft altitudes, and because the 205 $\mu$m transition frequency was poorly known.   The 205 $\mu$m line lies near the long wavelength cutoff for stressed Ga:Ge photoconductors, strongly limiting the sensitivity of spectrometers that employ these devices. However, the 122 $\mu$m [NII] line was first detected from the Kuiper Airborne Observatory in 1989 and the 205 $\mu$m line in 1991 from the Galactic HII region G333.6-0.2 (see Colgan et al. 1993, and references therein).  Since then, the 122 $\mu$m line was detected and mapped from a wide variety of sources using the Long Wavelength Spectrometer (LWS) on the Infrared Space Observatory (ISO).  The 205 $\mu$m line remains relatively unexplored, since the LWS has a long wavelength cutoff of 197 $\mu$m.  


The 205 $\mu$m [NII] line provides a vital key to the interpretation of the 158 $\mu$m [CII] line.  The [CII] line  is important.  It typically amounts to 0.3$\%$ of the observed far-IR luminosity for star-forming galaxies like the Milky Way (cf. Stacey et al. 1985, 1991; Crawford et al. 1985; Wright et al. 1991), and is the dominant coolant for much of the neutral interstellar medium (ISM) in galaxies including atomic clouds, the warm neutral medium, and photodissociation regions (PDRs) on the surfaces of molecular clouds exposed to stellar UV photons.  However, modeling of the neutral gas cooling is complicated by the fact that a significant fraction of the observed [CII] line emission may arise from low-density ionized gas regions.  Even for the Milky Way, it is uncertain how much of the observed [CII] line emission arises from low-density ionized gas, and how much arises from the neutral ISM.  Ionized gas phase contribution estimates range from 10 to 30$\%$ (Stacey et al. 1985) to near 50$\%$ (Petuchowski and Bennett 1993, Heiles 1994), depending on assumed gas-phase abundance of C$^{+}$ and ionized gas density.  Fortunately, the far-IR fine-structure lines of N$^{+}$ offer a method for unraveling the various contributions.  


The [NII] 205 $\mu$m line and the [CII] 158 $\mu$m line (n$_{crit} \sim$ 46 cm$^{-3}$, T$_{electron}$ = 8000 K) have nearly identical critical densities for excitation in ionized gas regions.  Therefore, the line ratio in ionized media is essentially only a function of the N$^{+}$/C$^{+}$ abundance ratio.  This is insensitive to the hardness of the stellar radiation fields since the photon energies required to ionize each species to the next ionization states are similar:  to form N$^{++}$ and C$^{++}$ requires 29.6 and 24.4 eV, respectively.  Therefore, the observed [NII] line emission reveals the fraction of the observed [CII] emission that arises from ionized gas, thereby better describing the cooling of both the ionized gas and the neutral ISM.  The [NII] observations, when combined with [CII] work, therefore constrain the cooling of two major components of the ISM.

Here we report the first detection of the 205 $\mu$m line from a ground-based observatory using a direct detection spectrometer\footnote{There is a near simultaneous detection of this line reported by Kawamura et al. (2006) made with a heterodyne reciever.}.  The line was detected from the radio peak of the Carina II HII region in the Carina Nebula using the South Pole Imaging Fabry-Perot Interferometer (SPIFI, Bradford et al. 2002) on the Antarctic Submillimeter Telescope and Remote Observatory (AST/RO) at South Pole (Stark et al. 2001).  We use these observations to probe the density structure of the nebula and to determine the fraction of the observed [CII] line emission that arises from the ionized medium.

\section{Observations}

The Carina molecular cloud complex is a series of star-forming molecular clouds that lie at a distance of about 2 kpc along the Carina spiral arm of the Galaxy.  In the optical band, the Carina Nebula (NGC 3372) is quite prominent, subtending about a square degree on the sky.  At the core of the region lie the bright HII regions Carina I and Carina II, illuminated respectively by the Trumpler 14 and Trumpler 16 star clusters.  These two clusters contain 36 O stars, including 6 of the 17 known O3 stars in the Galaxy, and 3 WN-A stars (see Davidson $\&$ Humphreys 1997, and references therein).  The most famous member of Trumpler 16 is the enigmatic, luminous blue variable star $\eta$ Carinae.  Eta Carinae is a single star with a mass in excess of 100 M$_{\odot}$;  it brightened by more than 5 stellar magnitudes for two decades in the mid-19$^{th}$ century.  With a current luminosity in excess of 10$^{6}$ L$_{\odot}$ (Cox et al. 1995), even at a distance of 2.3 kpc the dust-enshrouded star is the brightest extrasolar-system infrared source in the sky.  The cluster's extraordinary concentration of O stars is unique within the Galaxy and rivals the 30 Doradus region of the Large Magellanic Cloud in total luminosity.

We mapped the Carina I and II regions in the 205.178 $\mu$m [NII] line using SPIFI on the AST/RO telescope at South Pole.  SPIFI was deployed on AST/RO for the 2005 Austral winter and was operational from 3 August until 7 September.  A detailed analysis presenting all of the Carina Nebula [NII] data will appear in a future paper (Oberst et al., in prep.).  Here we present data obtained at the line peak, and at a position located $1.27\arcmin$ from the peak that is the nearest spatial position sampled by the Infrared Space Observatory (ISO) satellite.  The line peak data are a coaddition of 3 separate pointings obtained on 2005 August 23 and 24, while the ISO position data are a coaddition of 5 separate pointings obtained on 2005, August 20, 23, and 28.

SPIFI's detective elements are a $5\times5$ array of silicon bolometers held at 60 mK and arranged on a rectangular grid with a 65$\arcsec$ spacing between pixels.  The array is fed by an array of Winston cones yielding a 54$\arcsec$ beam (solid angle, $\Omega _{beam}$ = $5.4\times 10^{-8}$ sr). The beam size was roughly verified through scans of the limb of the Moon.  We employed a velocity resolution of 61 km s$^{-1}$ (FWHM) for these observations, and scanned 7.54 resolution elements, Nyquist sampled.  The velocity standard was set by observing the 205.4229 $\mu$m laser line used as the local oscillator source for the TREND laser (Yngvesson et al. 2004), also deployed on AST/RO for this season. 
We obtained 100 scans for each of three distinct pointings for a total of 12.3 hours integration time on the line peak position.  The three pointings were dithered by 2 pixels in the array to minimize flat-fielding errors.  At the ISO position, we obtained 447 scans in 5 distinct pointings totalling 18.3 hours of integration time.

Telescope pointing was initially achieved through raster scans of the Sun at low elevations in the 350 $\mu$m window during the Austral summer, refined in August through observations of the Moon, and finally verified by observations of the CO(7$\rightarrow$6) line in G333.6-0.2.  Final pointing accuracy was $\sim$ $1\arcmin$.  The array was flat fielded by rotating our chopper wheel at the entrance window of SPIFI to alternately sample between the chopper blade and the (very emissive) sky at 205 $\mu$m.  Line-of-sight transmission was obtained by performing skydips of the telescope at 205 $\mu$m on 5 September and linking the skydips to the NRAO 350 $\mu$m tipping radiometer using the scaling laws of Chamberlin et al.\ (2003). We estimate the 205 $\mu$m line-of-sight transmission towards the Carina Nebula was between 4.3$\%$ and 5.3$\%$ at the time of the observations.

The data were obtained by using a 3-position chop of the AST/RO telescope tertiary mirror with amplitude of 30$\arcmin$, thereby sampling sky emission at $\pm$ 30$\arcmin$ in azimuth with respect to the source position.  The data were calibrated by obtaining the system gain (mV signal/Kelvin) on hot and cold loads placed in the f-cone of the receiver, then correcting this gain by the known efficiency of the telescope at 200 $\mu$m ($\eta_{tel}$ = 51$\%$, Stark et al. 1997) and the measured transparency of the sky at the time of the observations.  

As calibrated by our chopper wheel, over the course of an hour integration, the best pixels in SPIFI achieve an instrument noise equivalent power (NEP) (referred to the front of the dewar) of $2.5 \times 10^{-15}$ W Hz$^{-1/2}$.  This is about a factor of 10 better than the best NEPs obtained with stressed Ge:Ga photoconductors at these wavelengths under similar backgrounds (cf. Colgan et al. 1993).  The sensitivity ratio is consistent with the much better detective quantum efficiency of bolometers ($\eta_{bol}\sim 50 \%$) compared with stressed Ge:Ga photoconductors at these wavelengths ($\eta_{Ge:Ga} \sim 3\%$, cf. Stacey et al. 1992) and the presence of generation-recombination noise in the photoconductors (cf. Rieke 2003). SPIFI's sensitivity corresponds to a double-sideband heterodyne receiver temperature of $\sim$ 150 K, or a factor of $\sim$ 7 better than the best reported values at 1.4 THz (Yngvesson et al. 2004).


\section{Results}

We have three independent pointings during which a pixel was pointed at the radio continuum peak of the Carina II nebula  (l  = 287.$^{\circ}$57,  b = -0.$^{\circ}$59;  RA =$10^{h}45^{m}01.^{s}0$, Dec = -59$^{\circ}$38$\arcmin$13$\arcsec$ (J2000), Retalack, 1983).  The spectra from these three pixels were individually corrected for telluric transmission and aperture efficiency, and calibrated as described above to yield units of T$_{MB}$.  The [NII] line is clearly detected with a statistical significance of 20 standard deviations (Figure 1).  The peak antenna temperature is 0.42 K, and the integrated line flux is 1.03 $\pm$ 0.05 $\times$ 10$^{-14}$ W m$^{-2}$ in our 54$\arcsec$ beam, giving an integrated line intensity of 60 K$\cdot$km s$^{-1}$.   Given SPIFI's velocity resolution of 61 km s$^{-1}$, the intrinsic linewidth is about 50 km s$^{-1}$.  The linewidth and the velocity centroid  (-35  km s$^{-1}$) are in good agreement with those of the radio recombination lines (Brooks, Storey, $\&$ Whiteoak 2001), but the line is somewhat broader and blueward of the neutral gas lines such as CI ($^{3}P_{1}\rightarrow^{3}P_{0}$) and CO(4$\rightarrow$3) (Zhang et al. 2001), consistent with an origin for the [NII] line in the ionized medium.

\section{Discussion}

The Carina Nebula is well-studied morphologically and spectroscopically from the X-ray to radio regime.  Most relevant to the present work is the far-infrared spectroscopic mapping obtained with the ISO satellite.  Mizutani, Onaka and Shibai (2002) obtained complete (43 to 197 $\mu$m) LWS low-resolving-power (R $\sim$ 200) spectra of the Carina Nebula.  A rectangular region was mapped from l = 287.$^{\circ}$0 to 287.$^{\circ}$65 and b = -0.$^{\circ}$78 to -0.$^{\circ}$23, which covers the prominent Carina I and Carina II HII regions. The LWS  beam size was $\sim$ 70 to 80$\arcsec$, but the raster scans were coarsely sampled so that data were obtained on a square grid with a spacing of 3$\arcmin$. The prominent far-IR fine-structure lines of [OIII], [OI], [CII], [NIII], and [NII] 122 $\mu$m are readily detected at nearly every position mapped.  The authors use the line ratios to build a model for the ionized gas regions in the nebula.  In particular, the ratio of the 52 to 88 $\mu$m [OIII] lines is sensitive to electron density for densities between 100 and 3000 cm$^{-3}$ (cf. Melnick et al. 1979).   They find two distinct components to the electron density:  a high-density (n$_{e}\sim 100-350 \; \rm{cm}^{-3}$) component at the Carina I and Carina II HII regions, and an extended low-density (n$_{e}\la 100\; \rm{cm}^{-3}$) component detectable over the entire $\sim$ 30 pc diameter region mapped.  Mizutani et al. (2004) find a correlation between the observed [CII] and [OI] 63 $\mu$m lines from which they estimate that 80$\%$ of the [CII] line emission comes from PDRs in the Carina Nebula as opposed to low-density ionized gas regions. 


\subsection{The [NII] Line Ratio}

The [NII] lines provide a distinct probe of the ionization structure and gas density in HII regions.  Formation of O$^{++}$ requires energetic 35.1 eV photons, so that the [OIII] lines trace HII regions formed by very early-type (O7) stars.  Lower energy photons (14.53 eV) can form N$^{+}$, so that these lines arise from HII regions formed by the softer UV radiation of late O-type or early B-type stars, or from the lower ionization "outskirts" of HII regions formed by earlier-type stars.  Since O$^{++}$ and N$^{++}$ have similar ionization potentials (35.1 and 29.6 eV, respectively), an O$^{++}$ zone contains mainly N$^{++}$ rather than N$^{+}$.  The twin [NII] lines have significantly lower critical densities (n$_{crit} \sim$ 44 cm$^{-3}$ and 293 cm$^{-3}$ (8000 K) for the 205 and 122 $\mu$m lines, respectively) for thermalization than the [OIII] lines (n$_{crit}\sim$ 510 cm$^{-3}$ and 3600 cm$^{-3}$ (8000 K) for the 88 and 52 $\mu$m lines, respectively), so that they make better probes of lower-density gas.

Unfortunately, the undersampled ISO/LWS mapping of Mizutani et al. (2002) does not directly sample the position where we obtained our spectrum displayed in Figure 1.  The nearest position is at l = 287.$^{\circ}$550, b = -0.$^{\circ}$583 (RA = 10$^{h}$44$^{m}$54$^{s}$.1, Dec =  -59$^{\circ}$37$\arcmin$17$\arcsec$ (J2000)), or about 1.$\arcmin$27 (more than a full beam) offset from our peak line flux position.  However, this position is covered by other pixels in the array, for which we obtain an integrated line flux of 21.4 K$\cdot$ km s$^{-1}$, or 3.6 $\pm$ 0.7 $\times$ 10$^{-15}$ W m$^{-2}$ in our 54$\arcsec$  beam.  Using the ISO data archive , we obtain a 122 $\mu$m [NII] line flux of 1.66 $\times 10^{-14}$ W m$^{-2}$, or making the extended source correction, a flux of 1.12 $\times 10^{-14}$ W m$^{-2}$ in the 78.2$\arcsec$ LW2 ISO/LWS detector beam (Gry et al. 2003).  Assuming a uniform intensity source, and correcting for the difference in beam size, the ISO flux is 5.33 $\times$ 10$^{-15}$ W m$^{-2}$ referred to our 54$\arcsec$ beam, so that the 122 $\mu$m line is 1.5 times brighter than the 205 $\mu$m line.  Assuming 30$\%$ calibration uncertainty for each line, the ratio is 1.5 $\pm$ 0.64.

We have calculated the [NII] 122 $\mu$m to 205 $\mu$m line intensity ratio as a function of gas density assuming electron impact excitation and using the collision strengths from Hudson and Bell (2004) scaled to an assumed electron temperature of 8000 K (Figure 2).  The observed ratio of 1.5 indicates a very modest gas density: $\sim$ 32 cm$^{-3}$.  Even allowing for a 30$\%$ calibration uncertainty, it is clear that the [NII] lines (tracing low ionization gas) arise from a low-density medium: 20 $\la n_{e} \la$ 60 cm$^{-3}$.  Therefore, the ``halo" of low density ionized gas discovered in its [OIII] line emission (Mizutani et al. 2002) also contains gas in lower ionization states.  This low ionization, diffuse gas is similar to the warm ionized medium in the Galaxy as a whole.

\subsection{The Fraction of [CII] from Ionized Gas}

Since the critical densities for electron impact excitation of the 158 $\mu$m [CII] line and the 205 $\mu$m [NII] line are very similar, to a good approximation the line ratio in ionized gas regions is dependent only on the relative abundance of C$^{+}$ and N$^{+}$ within the HII region.  Using the collision rates for exciting the ground-state levels of C$^{+}$ from Blum and Pradhan (1992), we have calculated the expected ratio of the two lines as a function of density and present it in Figure 2.  The temperature dependence is quite small, as the levels are only 91 and 70 K above ground, respectively---small compared with the temperature (8000 K) of an HII region.  For the calculation, we take the gas-phase abundances of n(C$^{+}$)/n$_{e} = 1.4 \times 10^{-4}$ and n(N$^{+}$)/n$_{e} = 7.9 \times 10^{-5}$ (Savage and Sembach 2004) and correct for the fraction of the two elements expected in the first ionization state using the HII region models of Rubin (1985).  The [CII]/[NII] line ratio from ionized gas is 3.1 at low densities, and 4.3 at high densities, and dips to below 3 between densities of 20 and 100 cm$^{-3}$ since the upper J=2 level of the N$^{+}$ begins to be significantly populated at densities above 20 cm$^{-3}$.

From the ISO archive, we obtain a [CII] flux of 1.01 $\times$ 10$^{-13}$ W m$^{-2}$.  Making the extended source correction, this corresponds to 5.5 $\times$ 10$^{-14}$ W m$^{-2}$ within the 70$\arcsec$ (Gry et al. 2003) LW4 beam. Again, assuming a uniform source the flux within our 54$\arcsec$ beam would be 3.3 $\times$ 10$^{-14}$ W m$^{-2}$.  Allowing for a 30$\%$ calibration uncertainty in each line, the [CII]/[NII] 205 $\mu$m line ratio is $\sim$ 9.2 $\pm$ 3.9 to 1.  Inspecting Figure 2 for a gas density of 32 cm$^{-3}$, if both lines arise from the ionized gas, we would expect the line ratio to be 2.44.  Therefore, 27$\%$ (with an upper bound of 46$\%$ and a lower bound of 19$\%$) of the observed [CII] flux arises from the ionized medium, and the remaining 73$\%$ must have its origin in the neutral ISM.  These results support and underpin prior work which contends that most of the observed [CII] line emission from Galactic star-forming regions, the Galaxy as a whole, and from external galaxies arises in warm dense photodissociation regions on the surfaces of molecular gas clouds.





\acknowledgments

This work was supported by NSF grants OPP-0094605, OPP-0338149, and OPP-0126090, and by NASA grant NNG05GK70H.  We are indebted to the Cardiff group under P. A. R. Ade for their excellent filters, to the GSFC group (Christine Allen, S. H. Moseley, D. J. Benford, and J. G. Staguhn) for their excellent bolometers, and to J. W. Kooi who set up local oscillators for our frequency calibration at Pole.  We are grateful to K. S. Yngvesson and the TREND group for use of the TREND laser, and for the ISO Archive for the online ISO-LWS data reduction tools.  We also thank the many people who have contributed to the success of SPIFI both at AST/RO and the JCMT including C. M. Bradford, A. D. Bolatto, J. A. Davidson, and M. L. Savage. Finally, we thank an annoymous referee for many helpful comments on an earlier draft of this manuscript.



\clearpage

\begin{figure}
\includegraphics[angle=0,scale=.50]{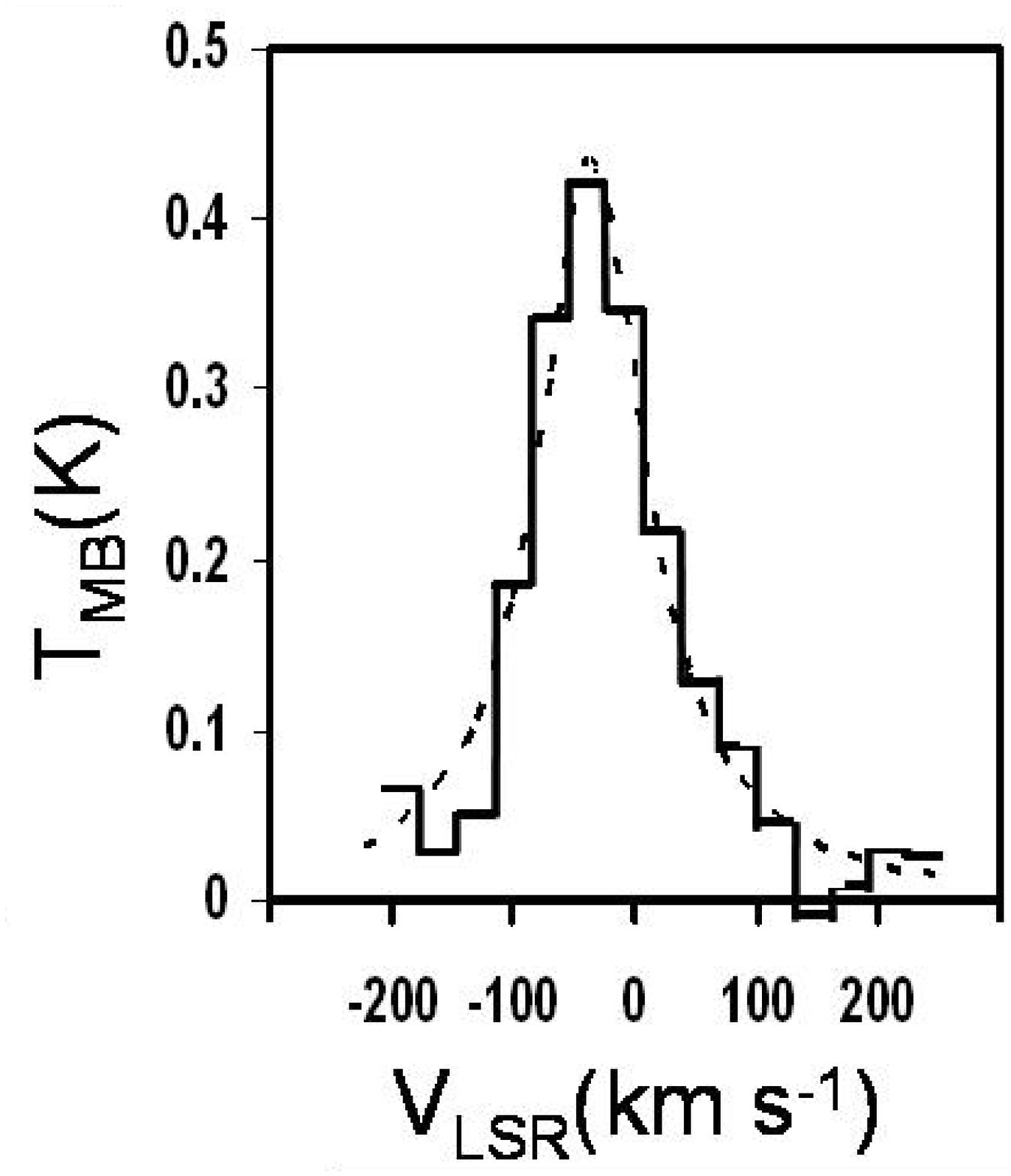}
\caption{[NII] 205 $\mu$m spectrum obtained using SPIFI on AST/RO.  These data were obtained at the peak of the radio continuum contours of the Carina II HII region (Retalack, 1983).  Velocity resolution is 60 km s$^{-}$1.  The  dashed line shows the intrinsic instrumental profile. The data were smoothed with a Hann window.\label{fig2}}
\end{figure}

\clearpage

\begin{figure}
\includegraphics[angle=0,scale=.50]{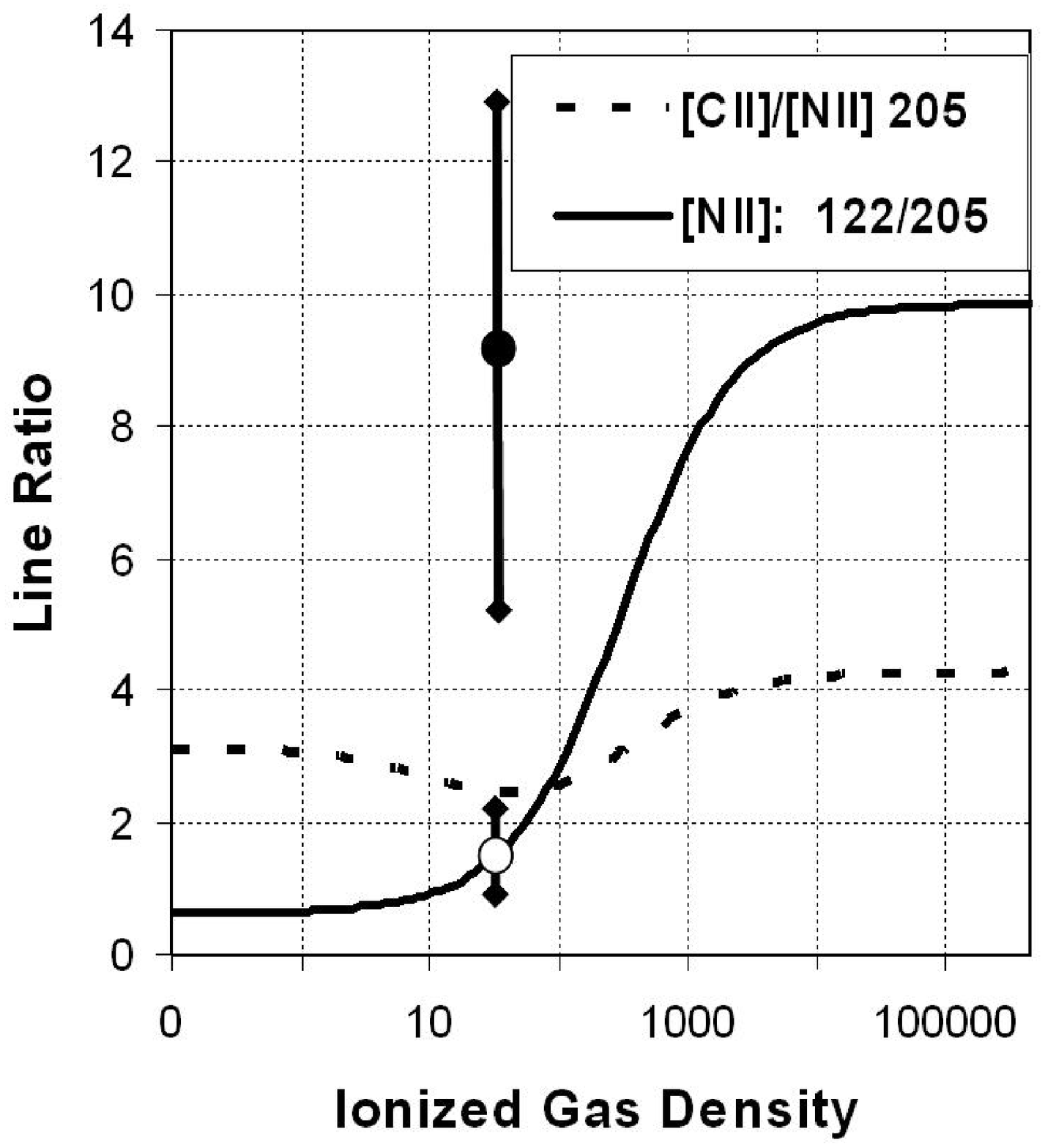}
\caption{[NII] 122 $\mu$m to 205 $\mu$m line intensity ratio as a function of ionized gas density, n$_{e}$. (solid line).  The open circle marks our observed line ratio of 1.5 $\pm$ 0.64.  The ratio of the [CII] 158 $\mu$m to [NII] 205 $\mu$m lines as a function of n$_{e}$ (dashed line). The filled circle marks the observed line ratio of 9.2 $\pm$ 3.9.}
\end{figure}


\begin{thebibliography}{}
\bibitem[]{221} Blum, R.D, $\&$ Pradhan, A.K. 1992 \apjs ~80, 425
\bibitem[]{222} Bradford, C.M., Stacey, G.J., Swain, M.R., Nikola, T., Bolatto, A.D., Jackson, J.M., Savage, M.L., Davidson, J.A., $\&$ Ade, P.A.R. 2002, Ap. Opt. 41, 2561
\bibitem[]{224}Brooks, K.J., Storey, J.W.V., $\&$ Whiteoak, J.B. 2001, \mnras ~327, 46
\bibitem[]{225}Chamberlin, R.A., Martin, R.N., Martin, C.L., $\&$ Stark, A.A. 2003, SPIE 4855, 609
\bibitem[]{226}Colgan, S.W.J., Haas, M.R., Erickson, E.F., Rubin, R.H., Simpson, J.P., $\&$ Russell, R. W. 1993, \apj ~413, 237
\bibitem[]{227}Cox, P., Mezger, P. G., Sievers, A., Najarro, F., Bronfman, L., Kreysa, E., \& Haslam, G. 1995 \aap ~297, 168
\bibitem[]{228}Crawford, M.K., Genzel, R., Townes, C.H., $\&$ Watson, D.M. 1985, \apj ~291, 755
\bibitem[]{229}Davidson, K., $\&$ Humphreys, R.M. 1997, \araa ~35, 1
\bibitem[]{231}Gry, C. et al. 2003, ESA SP-1262, 2003 
\bibitem[]{232}Heiles, C 1994, \apj ~436, 720
\bibitem[]{233}Hudson, C.E., $\&$ Bell. K.L. 2004, \mnras ~348, 1275
\bibitem[]{234}Kawamura et al. 2006 in prep
\bibitem[]{235}Kramer, C., Mookerjea, B., Bayer, E., Garcia-Burillo, S. Gerin, M., Israel, F.P., Stutzki, J., $\&$ Wouterloot, J.G.A. 2005, \aap ~441, 961
\bibitem[]{236}Melnick, G., Gull, G.E., $\&$ Harwit, M. 1979, \apjl ~227, L35
\bibitem[]{237}Mizutani, Onaka, $\&$ Shibai 2002, \aap ~382, 610 
\bibitem[]{238}Mizutani, Onaka, $\&$ Shibai 2004, \aap ~423, 579
\bibitem[]{239}Oberst, T.E. et al. 2006, in preparation
\bibitem[]{240}Petuchowski, S.J., $\&$ Bennett, C.L. 1993, \apj ~405, 591
\bibitem[]{241}Retallack, D.S. 1983, \mnras 204, ~669
\bibitem[]{242}Rieke, G. 2003, Detection of Light, 2nd ed.  Cambridge University Press
\bibitem[]{243}Rubin, R.H. 1985, \apjs ~57, 349
\bibitem[]{244}Savage, B.D., $\&$ Sembach, K.R. 1996, \araa ~34, 279
\bibitem[]{245}Stacey, G.J., Viscuso, P.J., Fuller, C.E., $\&$ Kurtz, N.T., 1985, \apj ~289, 803
\bibitem[]{246}Stacey, G.J., Geis, N., Genzel, R., Lugten, J.B., Poglitsch, A., Sternberg, A., $\&$ Townes, C.H.. 1991, \apj ~373, 423
\bibitem[]{247}Stacey, G.J., Beeman, J.W., Haller, E.E., Geis, N., Poglitsch, A., $\&$ Rumitz, M. 1992, IJMW 13, 1689 
\bibitem[]{248}Stark, A.A. et al. 2001, \pasp ~115, 567
\bibitem[]{249}Stark A.A., Chamberlin, R.A., Cheng, J., Ingalls, J., $\&$ Wright, G.,1997, Rev. Sci. Instrum., 68, 2200 
\bibitem[]{250}Wright, E.L. et al. 1991, \apj ~381, 200 
\bibitem[]{252}Yngvesson, S., et al. 2004, 15$^{th}$ Intern. Symp. Space THz Technology, eds.
G. Narayanan, 365
\bibitem[]{254}Zhang, X., Lee, Y., Bolatto, A.D., $\&$ Stark, A.A. \apj ~2001, 553, 274
\end{thebibliography}
\end{document}